\documentclass[11pt]{article}
\usepackage{graphicx}
\usepackage{amsmath}
\usepackage{amsfonts}
\usepackage{amssymb}
\usepackage{epsfig}
\usepackage{times}
\usepackage{cite}
\usepackage{calc}
\usepackage{version}
\usepackage[french,english]{babel}
\def\nDL{\line(2,1){12.5} \hspace{-0.38cm} {\rm DL}}

\begin{document}

%%%%%%%%%%%%%%%%%%%%%%%%%%%%%%%%%%%%%%%%%%%%%%%%%%%%%%%%%%%%%%%%%%%%%%%%%%%%%%

\begin{titlepage}

%%%%%%%%%%%%%%%%%%%%%%%%%%%%%%%%%%%%%%%%%%%%%%%%%%%%%%%%%%%%%%%%%%%%%%%%%%%%%%
%\today \hfill {\em draft version,  not for diffusion}

\null

\vskip 2.5cm

{\bf\large\baselineskip 20pt
\begin{center}
\begin{Large}
Medium-modified DGLAP evolution of fragmentation functions
from large to small $\boldsymbol{x}$
\end{Large}
\end{center}
}
\vskip 1cm

\begin{center}

S. Albino, B.A. Kniehl \& R. P\'erez-Ramos\\
\smallskip
{\em II. Institut f\"ur Theoretische Physik, Universit\"at Hamburg\\
Luruper Chaussee 149, D-22761 Hamburg, Germany}
\end{center}

\baselineskip=15pt

\vskip 3.5cm

{\bf Abstract}: The unified description of fragmentation function evolution from large to small $x$
which was developed for the vacuum in previous publications is now generalized to the medium,
and is studied for the case in which the complete contribution from the 
largest class of soft gluon logarithms, the double logarithms, are accounted
for and with the fixed order part calculated to leading order.
In this approach it proves possible to choose the remaining degrees of freedom related to the medium 
such that the distribution of produced hadrons
is suppressed at large momenta while the production of soft radiation-induced
charged hadrons at small momenta is enhanced, in agreement with experiment. 
Just as for the vacuum, our approach does not require further assumptions concerning fragmentation
and is more complete than previous computations of evolution in the medium.
   
\vskip 2 cm


\vfill

%%%%%%%%%%%%%%%%%%%%%%%%%%%%%%%%%%%%%%%%%%%%%%%%%%%%%%%%%%%%%%%%%%%%%%%%%%%%%

\end{titlepage}

%%%%%%%%%%%%%%%%%%%%%%%%%%%%%%%%%%%%%%%%%%%%%%%%%%%%%%%%%%%%%%%%%%%%%%%%%%%%%%
\section{Introduction}
The current description of the inclusive production of single hadrons in $e^+e^-$,
$ep$ and hadron collision experiments
is provided by the parton model of perturbative QCD (pQCD) involving vacuum fragmentation functions
(FFs) $D_a^h(x,Q^2)$ ($D$ is a vector containing all quark FFs $D_q$, all
antiquark FFs $D_{\bar q}$ and the gluon FF $D_g$), each of which corresponds at lowest order to the 
probability for a parton $a$ produced in the vacuum at short distance $1/Q$ to form a jet
that includes the hadron $h$ carrying a fraction $x$ of the longitudinal
momentum of $a$. Different theoretical approximations to the $Q^2$ evolution have been derived depending
on the kinematical region of $x$: fixed order (FO) calculations at intermediate
and large $x$ and resummation to all orders of soft gluon logarithms (SGLs)
at small $x$. In Ref.\ \cite{AKKPRL}, an approach which unifies the 
double logarithmic approximation (DLA) at small
$x$ and the leading order (LO) Dokshitzer-Gribov-Lipatov-Altarelli-Parisi 
(DGLAP) \cite{dglap} evolution of FFs at large $x$ was introduced.
This approach resums both SGLs and FO logarithms in a consistent
way and is defined to any order,
and thus allows for a determination of quark and gluon
FFs over a wider range of the data
than previously achieved. It was shown to describe well both the general
features of the ``hump-backed plateau" \cite{Basics} at small $x$, 
without affecting the quality of the excellent FO description of the large-$x$ region.

In this paper we are concerned with the more general case of medium-modified
FFs in heavy-ion collision experiments at all possible values of $x$. 
In contrast to fragmentation in a vacuum,
the fragmenting partons in these reactions are believed to first propagate through a quark-gluon plasma (QGP),
which affects the FF evolution. 
Indeed, recent experiments at the BNL Relativistic Heavy 
Ion Collider (RHIC) have revealed
the phenomenon of a strong suppression of hadrons at high transverse momentum 
\cite{PHENIX,STAR}, in contrast to the expectations of the parton model and the observations
from proton-(anti)proton collisions, which supports the picture 
that hard partons going through dense matter
suffer a significant energy loss prior to hadronization
in the vacuum (for a recent review see Ref.\ \cite{arleo}). Thus the 
study of medium-modified FFs may lead to a better understanding of the physical 
origin of the energy loss effect by serving as a QGP thermometer 
in the nuclear matter \cite{Armesto:2007dt}.
The study of parton energy loss and medium-modified observables
would ideally require the re-construction of jets in heavy-ion
collisions. Of course, the huge background makes this task highly delicate.
Nevertheless, thanks in particular to important theoretical
developments on the jet re-construction algorithms \cite{salam}
in a high-multiplicity environment, future analyses at the LHC 
by ALICE \cite{ALICE} and CMS \cite{CMS} look very promising.

Predictions for multiparticle production
in such reactions at small $x$ can be carried out by using a QCD-inspired model for the
effect of the medium on the single particle inclusive distribution 
inside high energy jets \cite{Basics}, the so-called 
``distorted hump-backed plateau". For example, such a model was applied in
Ref.\ \cite{Urs} in the framework of the 
modified leading logarithmic approximation (MLLA) . 
In this model, soft emissions, which manifest themselves in the DGLAP splitting
functions \cite{dglap} at small $x$ as SGLs in general and as
$1/x$ infra-red singularities at LO, are enhanced at LO
by multiplication of a nuclear medium factor $N_s>1$.
This enhancement can be interpreted as the phenomenological description
of the experimentally observed softening of jet spectra in nucleus-nucleus 
collisions. From the theoretical point of view, the enhancement can be seen as arising
from some effective lagrangian which would be suitable for processes 
in a dense nuclear environment. 

If $x$ is too small, the $1/x$ rise of the LO splitting functions is 
not physically correct since the FO approach becomes invalid. 
In physical applications, the effect of the enhancement on the complete
resummed contribution of double logarithms (DLs) to all orders, of which the $1/x$
term is only the LO contribution, must be calculated by applying this enhancement
to an evolution equation suitable for small $x$, 
such as the double logarithm approximation (DLA) that follows
from angular ordering (AO) \cite{Basics}, or the MLLA master equation as was done in Ref.\ \cite{Urs}.

In this paper, we extend the approach of Ref.\ \cite{AKKPRL} to the medium
in order to have a formalism which is suitable from large- to small-$x$ values.
This approach in the vacuum is more complete
than the MLLA since it also incorporates the FO part of the splitting functions, which
are important for describing the evolution at large $x$, and thus is more suitable for global fits.
In particular, this approach allows for the use of small-$x$ data to add further constraints on FFs at large $x$
to those already provided by large-$x$ data.
To apply this approach to the medium involves simply repeating the same steps but in the context of the medium: 
we resum the medium-modified DLs, being the largest SGLs,
in the standard DGLAP evolution by using the medium-modified DLA, in order to improve the small-$x$ 
description of medium-modified evolution,
while the effect of the medium on the FO part is imposed to ensure the correct description at large $x$.
Just as for the vacuum, this approach can be extended to higher orders in the FO part and to higher classes of SGLs.
Note that we do not impose momentum conservation, since partons can lose energy into the medium, 
which acts on the fragmenting partons as an external colour field.

The rest of the paper is organized as follows. 
In section \ref{MmodDGLAP}, we discuss the modifications to the FO splitting functions due to medium effects.
Using these results, we resum the DLs in section \ref{evollargetosmallx} using the DLA equation
to obtain a DGLAP evolution valid from large to small $x$.
Finally, in section \ref{section:conclusions} we present our conclusions.
The medium-modified form of the MLLA is derived in appendix \ref{MLLAlim}.

\section{Medium-modified fixed order DGLAP evolution equations}
\label{MmodDGLAP}

In both the vacuum and the medium, the DGLAP evolution takes the form
\begin{equation}\label{eq:dglapeq}
\frac{d}{d\ln Q^2}D(x,Q^2)=\int_x^1\frac{dz}{z}P(z,a_s(Q^2))
D\left(\frac{x}{z},Q^2\right),
\end{equation}
where $P$ is the matrix of splitting functions.
At large $x$, it is accurately calculated from the FO approach in perturbation theory,
which yields a truncation of the series
\begin{equation}\label{eq:expansionP}
P(x,a_s)=\sum_{n=1}^\infty a_s^n P^{(n-1)}(x)
\end{equation}
at some order in $a_s=\alpha_s/(2\pi)$, which has the
LO $Q^2$ evolution
$a_s(Q^2)=1/(\beta_0\ln(Q^2/\Lambda_{QCD}^2))$, where
$\beta_0=(11/6)C_A-(2/3)T_Rn_f$ is the first coefficient of
the beta function and $\Lambda_{QCD}$ is the asymptotic
scale parameter of QCD. 
In Mellin space, defined as the integral transformation
\begin{equation}\label{eq:mellin}
F(\omega)=\int_0^1dx\,x^{\omega}F(x)
\end{equation}
for any function $F(x)$, the convolution in
Eq.\ (\ref{eq:dglapeq}) becomes the simple product
\begin{equation}\label{eq:dglapmell}
\frac{d}{d\ln Q^2}D(\omega,Q^2)=P(\omega,a_s(Q^2))D(\omega,Q^2).
\end{equation}
The Mellin transform is invertible, with the inversion given by
\begin{equation}\label{eq:invmell}
F(x)=\frac1{2\pi i}\int_Cd\omega\,x^{-\omega-1}F(\omega),
\end{equation}
where the integration contour $C$ in the complex $\omega$ plane 
is parallel to the imaginary axis and to the right
of all singularities of the integrand $F(\omega)$.
By using charge conjugation and SU($n_f$) flavour symmetry to decompose the DGLAP equation 
into 3 simpler types of equations,
Eq.\ (\ref{eq:dglapmell}) can be solved analytically up to the desired order in $a_s$,
making the calculation of evolved FFs via the inverse Mellin transform
numerically more efficient than the direct integration of Eq.\ (\ref{eq:dglapeq}).
These 3 types of equations are the
DGLAP equations for the non-singlet and valence quark FFs in which $P$ is a 1$\times$1 matrix,
and the DGLAP equation for $D=(D_{\Sigma},D_g)$, where $P(x,a_s)$ is the 2$\times$2 matrix which we write as
\begin{eqnarray}
P=\left(\begin{array}{cc} P_{\Sigma \Sigma} & P_{\Sigma g} \\ P_{g\Sigma} & P_{gg}\end{array}\right).
\end{eqnarray}
We begin with a study of the medium modifications to the latter DGLAP equation.
After multiple rescattering in a dense nuclear medium, 
a relativistic parton loses a significant fraction of its
energy scale. In the soft multiple momentum transfer model of the medium \cite{Baier:1996sk}, 
the singular terms $\propto1/x$ (as $x\to0$) and, by the LO symmetry of the parton splitting process for all $x\neq 1$ under
$x\rightarrow 1-x$, also the singular terms
$\propto1/(1-x)$ (as $x\to1$) are simply multiplied by a nuclear factor $N_s$ \cite{Urs} such that
soft emission is enhanced.
Applying these modifications to the LO splitting functions in the vacuum \cite{Basics,dglap},
the LO splitting functions in the medium using the notation of Eq.\ (\ref{eq:expansionP}) become
\begin{eqnarray}\label{eq:medsplitfunt}
P_{\Sigma g}^{(0)}(x)\!\!&\!\!=\!\!&\!\!4C_F\left(\frac{N_s}{x}+f_{\Sigma g}(x)\right),\quad 
P_{\Sigma \Sigma}^{(0)}(x)=2C_F\left(\left[\frac{N_s}{1-x}\right]_+ +f_{\Sigma \Sigma}(x)\right),\\
P_{gg}^{(0)}(x)\!\!&\!\!=\!\!&\!\!2C_A\left(\frac{N_s}{x}+\left[\frac{N_s}{1-x}\right]_+
+f_{gg}(x)\right),\quad
P_{g\Sigma}^{(0)}(x)=n_fT_Rf_{g\Sigma}(x),
\label{eq:medsplitfunt1}
\end{eqnarray}
where $C_F$ and $C_A$ are respectively the casimirs of the fundamental and adjoint
representations of the QCD color gauge group $SU(3)_c$, $T_R=1/2$ and $n_f$ is the
number of active (anti)quark flavours. 
The plus distribution applied to a function $F(x)$, written $[F(x)]_+$, is defined by
\begin{equation}
\int_0^1 dx [F(x)]_+ g(x)=\int_0^1 dx F(x) \left[g(x)-g(1)\right]
\end{equation}
for any function $g(x)$.
The functions $f_{ab}(x)$ are regular in the limit
$x\to0$. In particular, the Borghini-Wiedemann
choice \cite{Urs} is recovered by choosing the $f_{ab}(x)$ to be equal to their vacuum forms 
$\left[f_{ab}(x)\right]_{N_s=1}$ given for $x\neq 1$ by
\begin{subequations}
\begin{eqnarray}
\left[f_{\Sigma g}(x)\right]_{N_s=1}\!\!&\!\!=\!\!&\!\!\frac{x}2-1,\quad 
\left[f_{\Sigma \Sigma}(x)\right]_{N_s=1}=-\frac{x}2-\frac12,\\
\left[f_{gg}(x)\right]_{N_s=1}\!\!&\!\!=\!\!&\!\!x(1-x)-2,
\quad \left[f_{g\Sigma}(x)\right]_{N_s=1}=x^2+(1-x)^2.
\end{eqnarray}
\end{subequations}
(The delta functions at $x=1$ along the diagonal of the splitting functions can be determined by momentum conservation.)

However, the correct model for the medium is not known at present, therefore neither the value of $N_s$ nor
the $f_{ab}(x)$ are known.
We note that a more complete form for the medium-modified splitting functions $P(x,a_s)$
may be derived from some effective lagrangian ${\cal L}_{med}\propto (F^c_{\mu\nu,med})^2$
in the medium.

The DGLAP evolution of the non-singlet and valence quark FFs is accurately described for large- to small-$x$ values 
when the splitting function is calculated in the FO approach, since it is free of SGLs to all orders.
Furthermore, the splitting function behaves like $\sim 1/(1-x)$ to all orders.
Therefore, the effect of the medium on the large-$x$ LO evolution of these FFs is incorporated
simply by multiplying the $1/(1-x)$ singularities in the non-singlet and valence quark splitting functions by $N_s$,
and replacing the rest of these functions by an unknown function which is non singular as $x\rightarrow 0,1$ 
(apart from possible delta functions at $x=1$).

In Mellin space,
\begin{eqnarray}
P_{\Sigma g}^{(0)}(\omega)\!\!&\!\!=\!\!&\!\!4C_F\left(\frac{N_s}
{\omega}+f_{\Sigma g}(\omega)\right),\label{eq:p0qg}\\
P_{\Sigma \Sigma}^{(0)}(\omega)\!\!&\!\!=\!\!&\!\!2C_F\left( -N_s S_1(\omega)+
f_{\Sigma \Sigma}(\omega)\right),\label{eq:p0qq}\\
P_{gg}^{(0)}(\omega)\!\!&\!\!=\!\!&\!\!2C_A \left(\frac{N_s}{\omega}-N_s S_1(\omega)
+f_{gg}(\omega)\right)\label{eq:p0gg},\\
P_{g\Sigma}^{(0)}(\omega)\!\!&\!\!=\!\!&\!\!n_fT_Rf_{g\Sigma}(\omega),\label{eq:p0gq}
\end{eqnarray}
where the harmonic sum $S_1(\omega)=\sum_{n=1}^\omega 1/n$ for integer $\omega$.
The Mellin transform of the small-$x$ singularity $1/x$ is proportional to the 
small-$\omega$ singularity $1/\omega$ (note $S_1(0)=0$),
while the Mellin transform of the large-$x$ singularity $1/(1-x)$ grows at large $|\omega|$ as $\ln \omega$ 
(because $S_1(\omega)\simeq \ln \omega$ here).
Because $P_{\Sigma g}^{(0)}(\omega)$ 
and $P_{g\Sigma}^{(0)}(\omega)$ remain finite as $|\omega|$ approaches infinity, while
\begin{equation}
P_{aa}^{(0)}(\omega)
\approx-2C_aN_s\ln \omega,\qquad \omega\gg1,
\end{equation}
where
\begin{equation}
C_\Sigma =C_F,\,\,C_g=C_A,
\end{equation}
one finds, after using Eqs.\ (\ref{eq:dglapmell}) and (\ref{eq:invmell}), 
the following behavior of the quark singlet and gluon FFs at large $Q$:
\begin{equation}\label{eq:largexD}
D_a (x,Q^2)\sim(1-x)^{-1+4C_aN_s\Delta\xi(a_s(Q^2))},\quad 
\Delta\xi(a_s)=\frac1{2\beta_0}\ln a_s.
\end{equation}
The non-singlet and valence quark FFs will exhibit the same behaviour with $C_a=C_F$.
Therefore, FFs at large $x$ decrease when $N_s$ increases (provided that the exponent in Eq.\ (\ref{eq:largexD})
is positive, which is the case at large $Q$). 
Consequently, the production 
of hard hadrons gets {\it restricted} at large $x$ or, equivalently, in the
large-$p_T$ region in particle production in heavy-ion collisions.

The small-$x$ behavior of FFs
is controlled by the behavior of the splitting functions in Mellin space
near $\omega=0$.
In the next section we will show that,
in this region, the approximate behavior of $D_a(x,Q^2)$
as a function of $\ell=\ln(1/x)$ is given by
\begin{equation}\label{eq:smallxD}
xD_a(x,Q^2)\sim {\cal N}\exp\left[-\frac1{2\sigma^2}(\ell-\bar\ell)^2\right],
\end{equation}
where the average multiplicity increase is given by \cite{Redmed}
\begin{equation}\label{eq:dlamult}
{\cal N}\sim\exp\left(\frac1{\beta_0}
\sqrt{\frac{4N_sC_A}{a_s(Q^2)}}\right).
\end{equation}
Hence
the production of soft hadrons gets {\it enhanced}
in the medium at small $x$.

\section{Evolution equations from large to small $\boldsymbol{x}$}
\label{evollargetosmallx}

In the medium, the DLA-improved evolution equation for $D=(D_\Sigma,D_g)$, which was introduced for the vacuum in Ref.\ \cite{AKKPRL}
and which is valid from large to small $x$, becomes
\begin{eqnarray}\label{DLAeq}
\frac{d}{d\ln Q^2}D(z,Q^2)\!\!&\!\!=\!\!&\!\!\int_z^1\frac{dz'}{z'}
\frac{2C_AN_s}{z'}Az^{'2\frac{d}{d\ln Q^2}}\left[a_s(Q^2)
D\left(\frac{z}{z'},Q^2\right)\right]\notag\\
\!\!&\!\!+\!\!&\!\!\int_z^1\frac{dz'}{z'}
\bar P(z',a_s(Q^2))D\left(\frac{z}{z'},Q^2\right).
\label{eq:SGFO}
\end{eqnarray}
where the matrix $A$ is given by
\begin{equation*}
{A} =\left(
\begin{array}{cc}
        0 &\frac{2C_F}{C_A}  \cr
        0 & 1 
\end{array}\right).
\end{equation*}
Equation (\ref{DLAeq}) is obtained by multiplying by $N_s$ the $1/x$ factor in the first term on the right hand side.
At LO, the $1/(1-x)$ singularities in the function $\bar P$ must be similarly enhanced,
as dictated by the soft multiple momentum transfer model of the medium.
Setting $N_s=1$ in Eq.\ (\ref{eq:SGFO}), 
we recover the DLA-improved vacuum evolution equation of Ref.\ \cite{AKKPRL}.
Since $\bar P$ is free of DLs to all orders, Eq.\ (\ref{eq:SGFO}) can be used to determine the
complete DL contribution to $P$.
We note that $\bar P$ contains other SGLs beyond LO, which are less significant than DLs but significant nonetheless,
and the effect of the medium on these SGLs is not known.

By using the approach of Ref.\ \cite{AKKPRL},
we will use Eq.\ (\ref{eq:SGFO}) to obtain an evolution which is valid to DL accuracy at small $x$ and to LO accuracy at large $x$.
This goal can be achieved by repeating precisely the steps of Ref.\ \cite{AKKPRL}, which we briefly outline here:
We use Eq.\ (\ref{eq:mellin}) to put Eq.\ (\ref{eq:SGFO}) in Mellin space,
\begin{equation}\label{eq:mellineq}
\left(\omega+2\frac{d}{d\ln Q^2}\right)
\frac{d}{d\ln Q^2}D=2C_AN_sa_sAD+
\left(\omega+2\frac{d}{d\ln Q^2}\right)\bar P D,
\end{equation}
where we have set $D= D(\omega,Q^2)$, $a_s=a_s(Q^2)$ and $\bar P = \bar P(\omega,a_s)$ for brevity.
Substituting Eq.\ (\ref{eq:dglapmell}) into Eq.\ (\ref{eq:mellineq})
and setting 
\begin{equation}
P=P^{\rm DL}+P^{\nDL},
\label{DLdecofP}
\end{equation}
where $P^{\rm DL}$ represents the complete DL contribution to $P$ 
and the remainder $P^{\nDL}$ everything else, and then taking the leading term (see Refs.\ \cite{AKKPRL,Albino:2008gy} for a
more detailed discussion),
one gets the equation
\begin{equation}
2\left(P^{\text{DL}}\right)^2+\omega P^{\text{DL}}-2C_AN_sa_sA=0,
\end{equation}
whose only solution consistent with perturbation theory is \cite{AKKPRL}
\begin{equation}
P^{\text{DL}}(\omega,a_s)=\frac{A}{4}
\left(-\omega+\sqrt{\omega^2+16C_AN_sa_s}\right),
\end{equation}
which is the Mellin transform of
\begin{equation}\label{PDLinxspace}
P^{\text{DL}}(x,a_s)=\frac{A\sqrt{C_AN_sa_s}}{x\ln\frac1x}
J_1\left(4\sqrt{C_AN_sa_s}\ln\frac1x\right),
\end{equation}
where $J_1$ is the Bessel function of the first kind.

Our unified approach to medium-modified evolution 
in the case of DL accuracy at small $x$ and LO accuracy at large $x$
can now be formulated, as follows:
The evolution is performed using Eq.\ (\ref{eq:dglapeq}) with $P$ approximated as in Eq.\ (\ref{DLdecofP}),
where $P^{\rm DL}$ is given by Eq.\ (\ref{PDLinxspace}) and $P^{\nDL}$ is set equal to the
LO splitting function after its DLs have been subtracted to prevent double counting, viz.\
\begin{equation}
P(\omega,a_s)=P^{\rm DL}(\omega,a_s)+a_s P^{\nDL (0)}(\omega),
\label{DLdecofP2}
\end{equation}
\begin{figure}[h!]
\parbox{.49\linewidth}{
\begin{center}
\includegraphics[width=6.7cm,angle=-90]{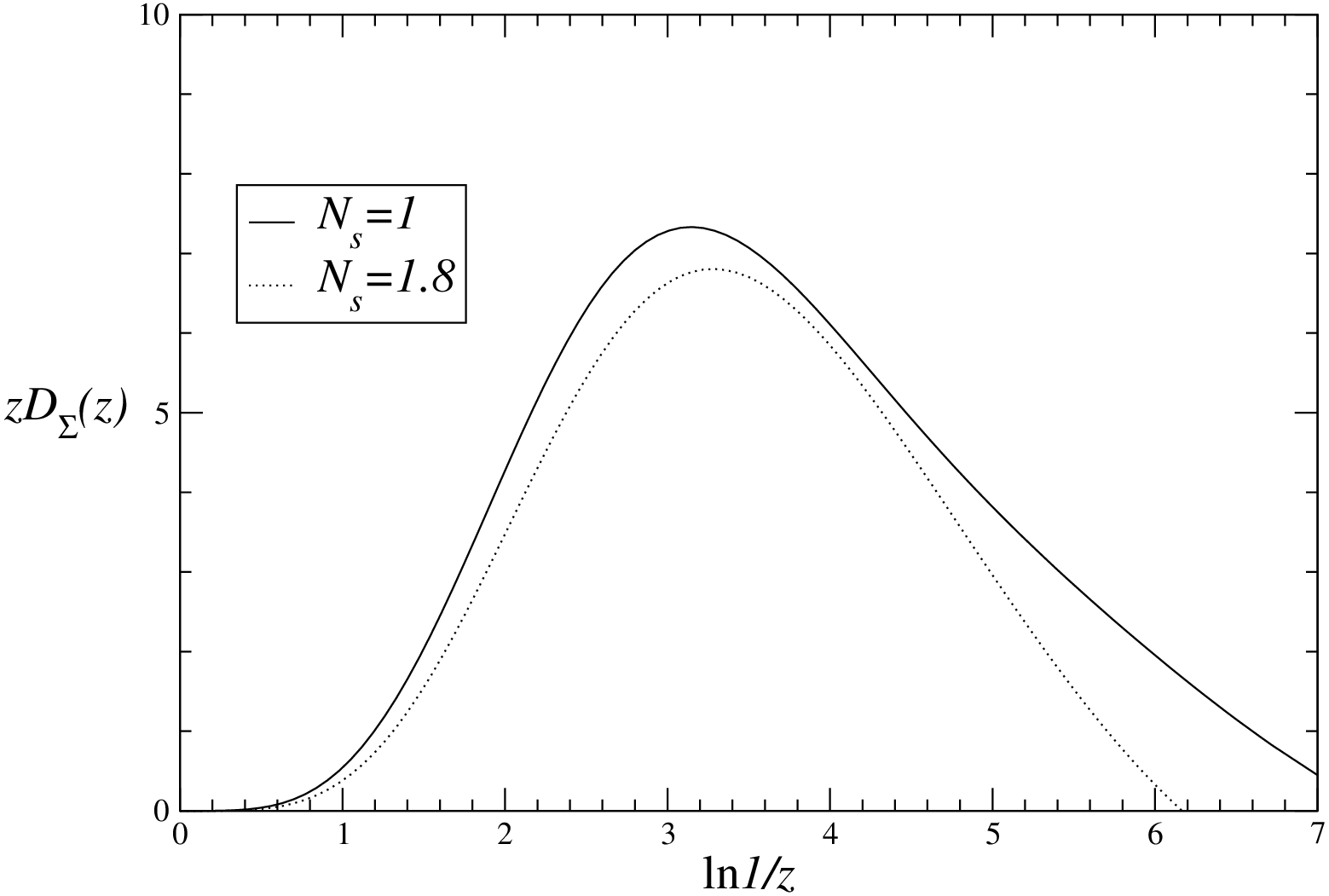}
\end{center}
}\parbox{.49\linewidth}{
\begin{center}
\includegraphics[width=6.7cm,angle=-90]{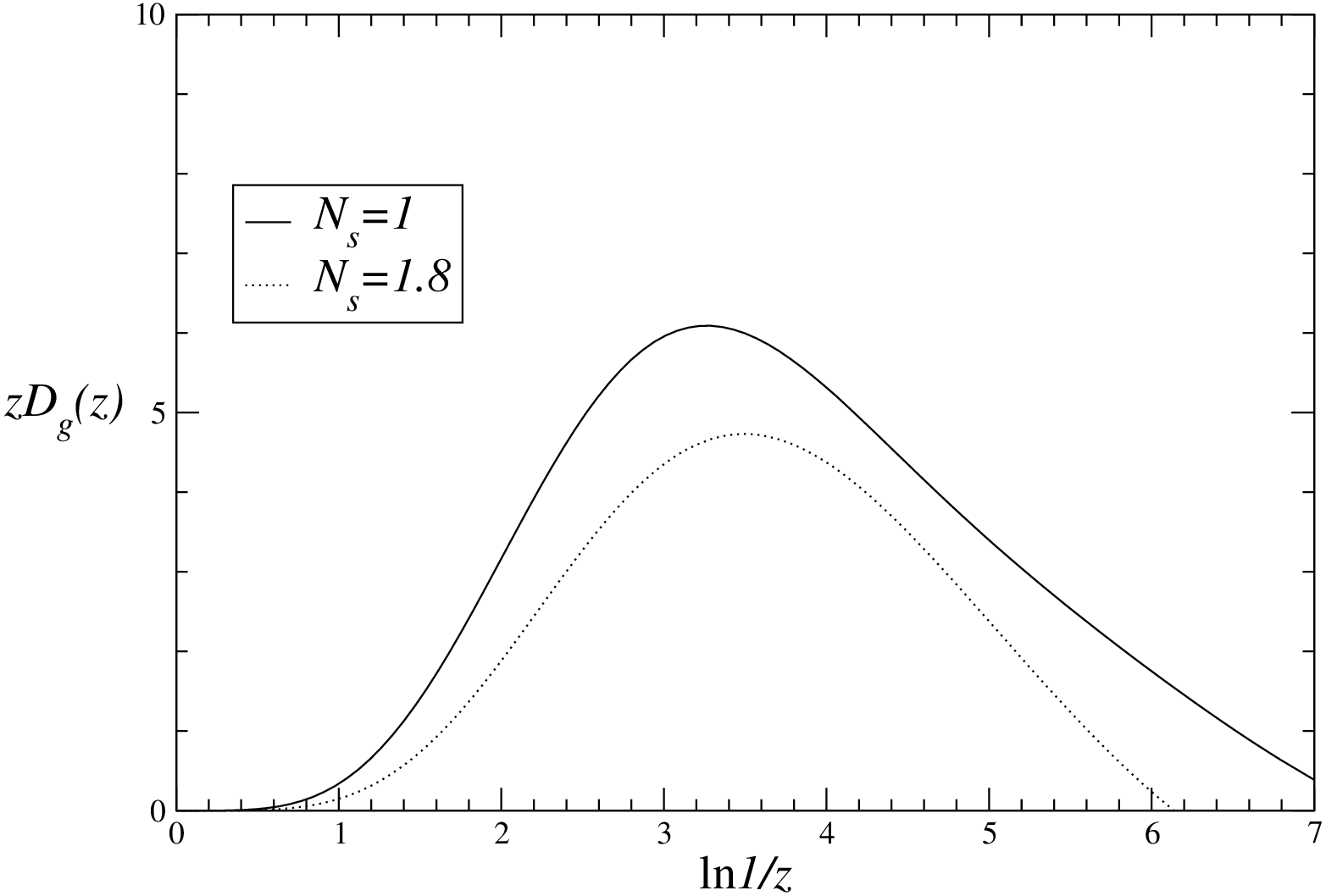}
\end{center}}
\caption{Quark singlet (left) and gluon (right) FFs evolved in the vacuum ($N_s=1$) and in the medium ($N_s=1.8$)
from a factorization scale $Q=10$ to $100$ GeV and with the choice $f_{ab}=N_s \left[f_{ab}\right]_{N_s=1}$. At the initial scale,
the choice $D_g=(C_A/(2C_F))D_\Sigma =N \exp[-c\ln^2 z] x^\alpha (1-x)^\beta$ is made where,
motivated by the results of Ref.\ \cite{AKKPRL}, we have taken
$N=1.6$, $c=0.35$, $\alpha=-2.63$ and $\beta=3$.
\label{QGplot}}
\end{figure}
\begin{figure}[h!]
\parbox{.49\linewidth}{
\begin{center}
\includegraphics[width=6.7cm,angle=-90]{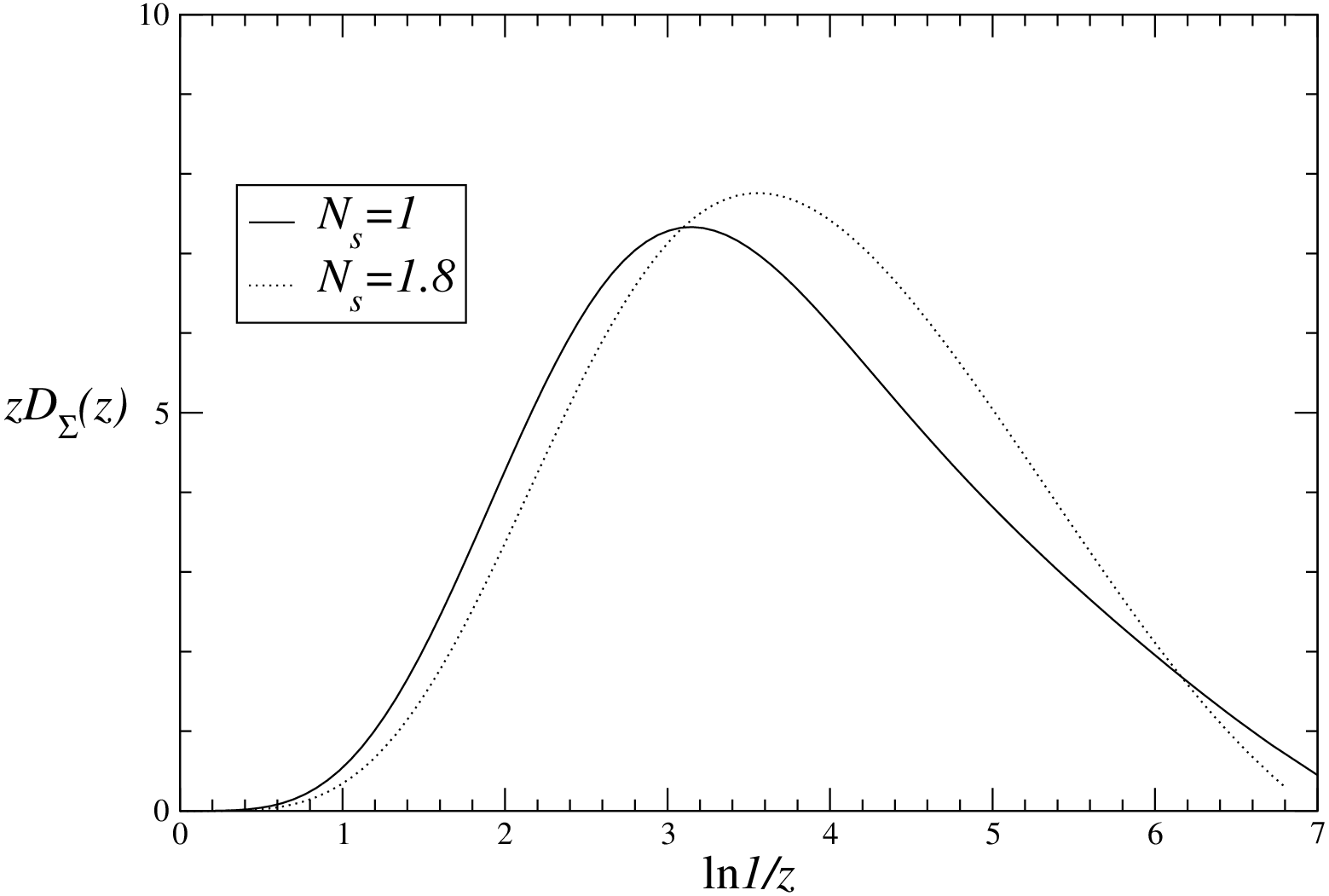}
\end{center}
}\parbox{.49\linewidth}{
\begin{center}
\includegraphics[width=6.7cm,angle=-90]{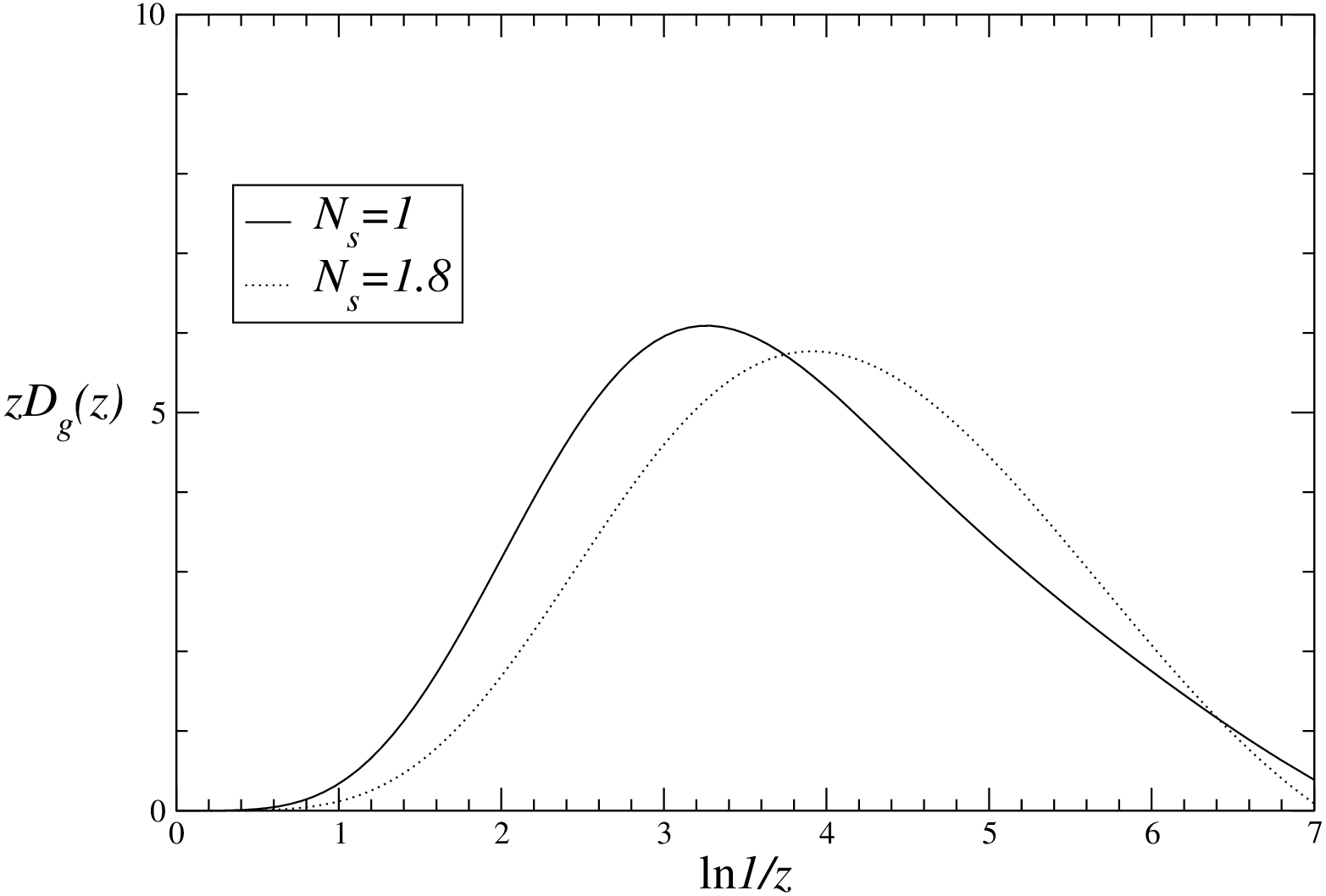}
\end{center}}
\caption{As in Fig.\ \ref{QGplot}, but with $f_{ab}=\left[f_{ab}\right]_{N_s=1}$.
\label{QGplot2}}
\end{figure}
\begin{figure}[h!]
\parbox{.49\linewidth}{
\begin{center}
\includegraphics[width=6.7cm,angle=-90]{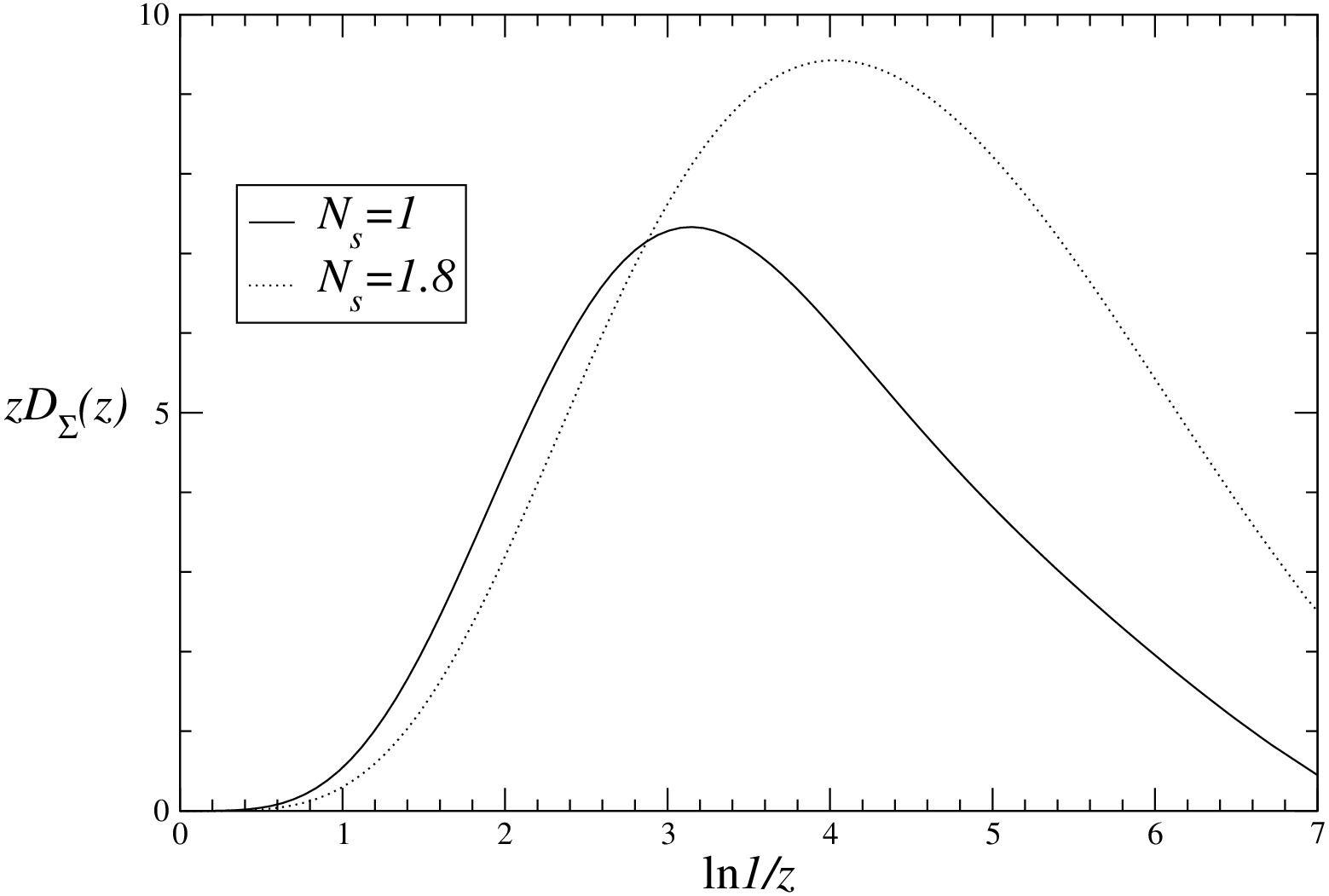}
\end{center}
}\parbox{.49\linewidth}{
\begin{center}
\includegraphics[width=6.7cm,angle=-90]{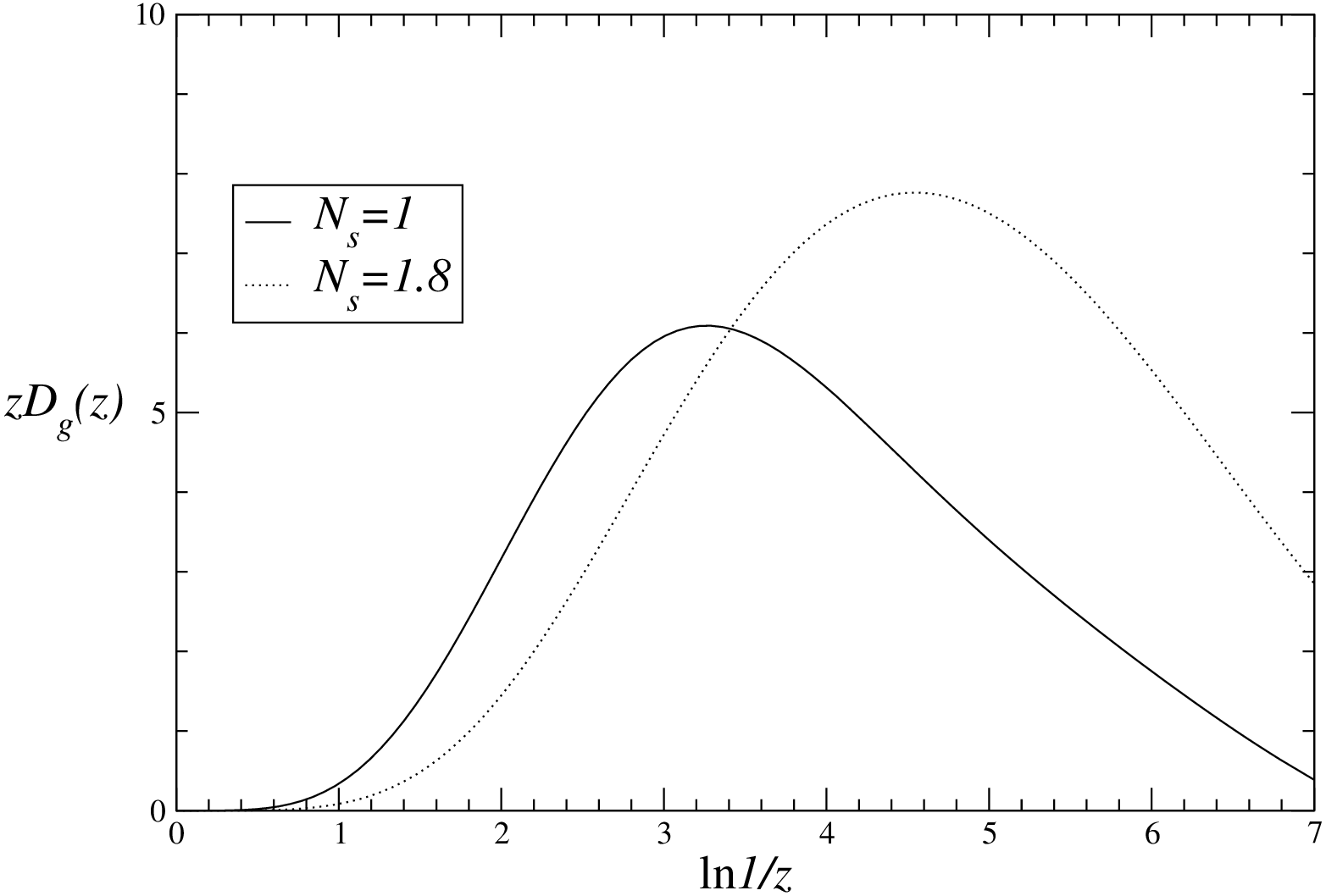}
\end{center}}
\caption{As in Fig.\ \ref{QGplot}, but with $f_{ab}=0$.
\label{QGplot3}}
\end{figure}
\begin{figure}[h!]
\parbox{.49\linewidth}{
\begin{center}
\includegraphics[width=6.7cm,angle=-90]{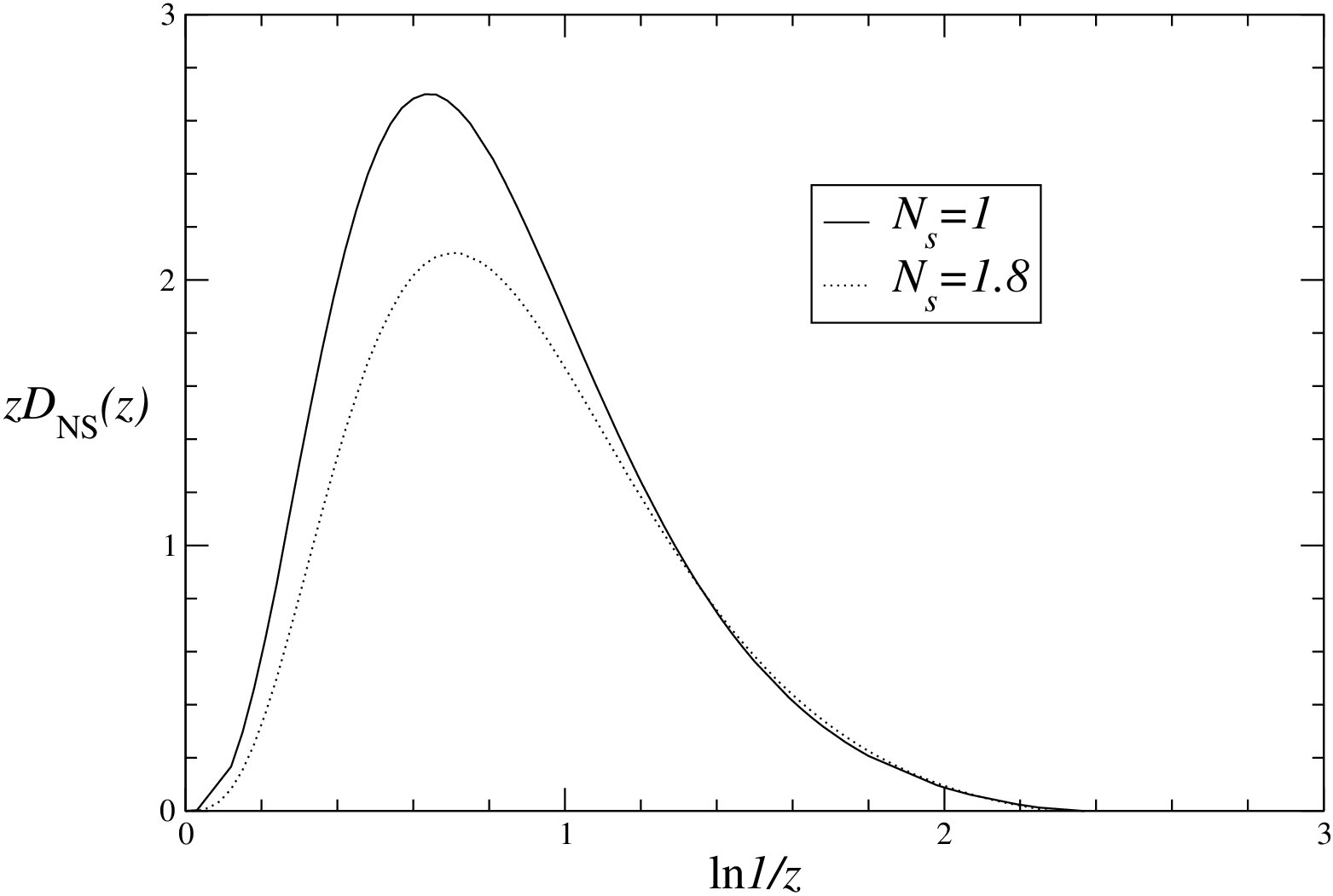}
\end{center}
}\parbox{.49\linewidth}{
\begin{center}
\includegraphics[width=6.7cm,angle=-90]{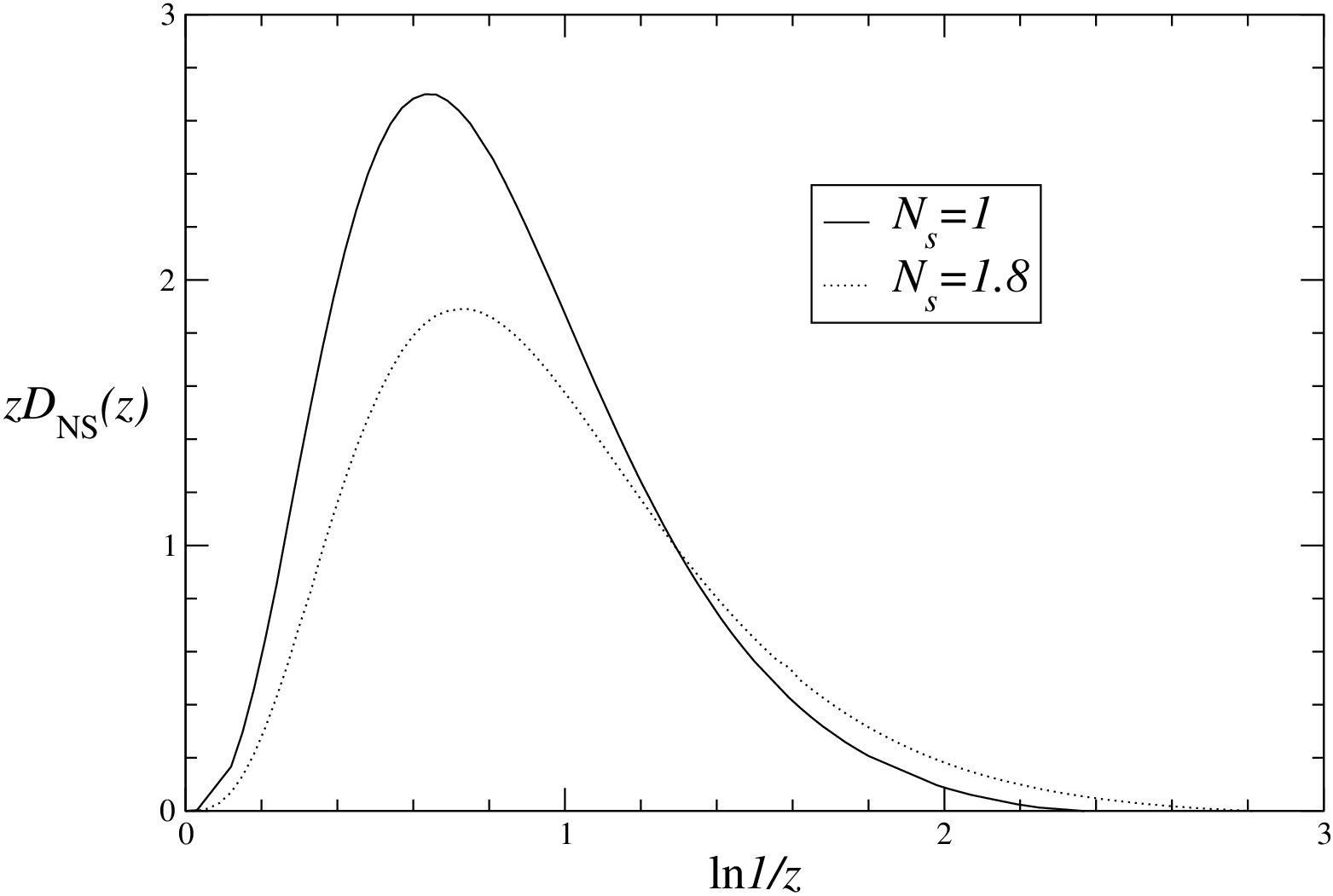}
\end{center}}
\begin{center}
\includegraphics[width=6.7cm,angle=-90]{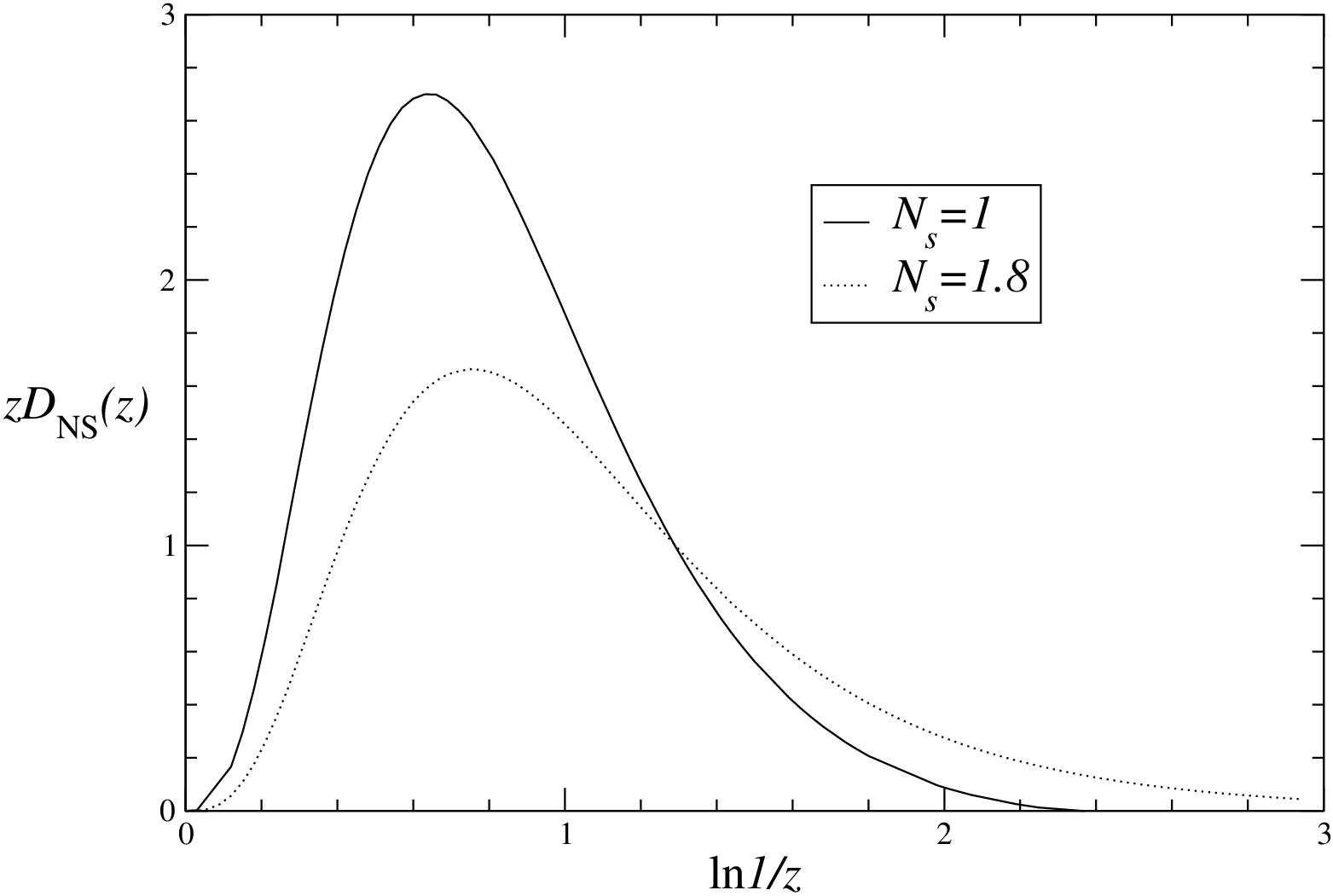}
\end{center}
\caption{As in Figs.\ \ref{QGplot} (top left), \ref{QGplot2} (top right) and \ref{QGplot3} (bottom), but for the non-singlet quark FF
with $N=500$, $c=0$, $\alpha=3$ and $\beta=3$.
\label{QGplot4}}
\end{figure}
where $a_s P^{\nDL (0)}(0)$ is the LO term in the expansion of $P^{\nDL}(\omega,a_s)$ in $a_s$.
Higher order terms in $P^{\nDL}(\omega,a_s)$ may also be included, 
provided the SGLs contained within them are subtracted \cite{AKKPRL} (or resummed, which is not possible at present).
In other words, the splitting functions in Mellin space are given by
\begin{eqnarray}\label{Pinmedium}
P_{\Sigma g}^{(0)}(\omega)\!\!&\!\!=\!\!&\!\!
\frac{C_F}{2C_A}
\left(-\omega+\sqrt{\omega^2+16C_AN_sa_s}\right)
+4C_Ff_{\Sigma g}(\omega),\nonumber \\
P_{\Sigma \Sigma}^{(0)}(\omega)\!\!&\!\!=\!\!&\!\!-2C_FN_sS_1(\omega)+
2C_Ff_{\Sigma \Sigma}(\omega),\nonumber \\
P_{gg}^{(0)}(\omega)\!\!&\!\!=\!\!&\!\!
\frac{1}{4}\left(-\omega+\sqrt{\omega^2+16C_AN_sa_s}\right)-2C_AN_s S_1(\omega)
+2C_Af_{gg}(\omega),\nonumber \\
P_{g\Sigma}^{(0)}(\omega)\!\!&\!\!=\!\!&\!\!n_fT_Rf_{g\Sigma}(\omega).
\end{eqnarray}
The results in Eq.\ (\ref{Pinmedium}) are the main results of this paper.

In order to understand better the $f_{ab}$, we plot the evolved FFs in the vacuum and in the medium
for different choices of them to see which give the most physically expected results, namely
an enhancement of hadron production at small $x$ due to induced soft gluon radiation by the medium, and a suppression
of this production at large $x$ due to parton energy loss in the medium.
We choose a fragmentation scale of 100 GeV which is suitable for the LHC.
To account for medium effects, we adopt the value $N_s=1.8$ from Ref.\ \cite{Urs}.
We choose the vacuum and medium FFs to be equal at 10 GeV, which is consistent with a finite-sized medium.
In Fig.\ \ref{QGplot}, we choose $f_{ab}=N_s \left[f_{ab}\right]_{N_s=1}$ for the evolution in the medium
(we absorb the delta functions at $x=1$ into the definition of the $f_{ab}$ here and in what follows),
i.e.\ the magnitudes of the $f_{ab}$ are chosen to be greater than their vacuum values.
In Fig.\ \ref{QGplot2}, we choose the vacuum values exactly, $f_{ab}=\left[f_{ab}\right]_{N_s=1}$,
while in Fig.\ \ref{QGplot3} we set $f_{ab}=0$.
Therefore, physically expected results can be obtained when the $f_{ab}$ are chosen to not increase with $N_s$.
Note that for all choices of the $f_{ab}$ there is always a suppression at large $x$.

Finally, we show the non-singlet (or valence) quark evolution in Fig.\ \ref{QGplot4}.
Such an FF gives a measure of the excess of e.g.\ charged over neutral kaon production \cite{Christova:2008te}, 
or of positively charged over negatively charged hadrons.
In this case there are no soft gluon logarithms at small $x$, and therefore there is no Gaussian behaviour at small $x$,
but rather an $x^\alpha$ behaviour, and the FO approach for the evolution is expected to be reliable from the 
largest- to smallest-$x$ values.
Just as for the quark singlet and gluon FFs, we find that the medium induces 
a suppression at large $x$ and an enhancement at small $x$ provided
that the $f_{ab}$ do not increase with $N_s$.

\subsection{Small-$\boldsymbol{x}$ limit \label{smallx}}

We now derive the small-$x$ behaviour given by Eqs.\ (\ref{eq:smallxD}) and (\ref{eq:dlamult}).
The medium modifications to the MLLA will be derived in appendix \ref{MLLAlim} as an extension to the results below.
In the small-$x$ region we may neglect the second term in Eq.\ (\ref{DLdecofP2}),
i.e.\ we may take
\begin{equation}
P(\omega,a_s)=P^{\rm DL}(\omega,a_s).
\end{equation}
Then the solution to the gluon component of Eq.\ (\ref{eq:gjet}) reduces to the simple form
\begin{equation}\label{eq:solDg}
D_g(\omega,Q^2)=E_{gg}(\omega,a_s(Q^2),a_s(Q_0^2))
D_g(\omega,Q_0^2)
\end{equation}
with
\begin{equation}\label{eq:solDgbis}
E_{gg}(\omega,a_s(Q^2),a_s(Q_0^2))=
\exp\left[\int_{Q_0}^Qd\ln Q'^2P_{gg}(\omega,a_s(Q'^2))\right],
\end{equation}
where 
\begin{eqnarray}\label{eq:anomdim}
P_{gg}(\omega,a_s)\!\!&\!\!=\!\!&\!\!\frac14\left(-\omega+\sqrt{\omega^2+16C_AN_sa_s}\right).
\end{eqnarray}
At small $x$, one can predict the shape and normalization
of the medium-modified FFs using the same procedure as for 
the vacuum FFs in Ref.\ \cite{fonweb}, where the expression
for $\left[D_a(x,Q^2)\right]_{N_s=1}$ was derived by writing the solution
in the form of a distorted Gaussian,
\begin{equation}\label{dg}
xD_g(x,Q^2)=\frac{{\cal N}}{\sigma\sqrt{2\pi}}
\exp\left[\frac18k-\frac12s\delta-\frac14(2+k)\delta^2
+\frac16s\delta^3+\frac1{24}k\delta^4\right].
\end{equation}
In this expression, $\delta=(\ell-\bar\ell)/\sigma$, ${\cal N}$ is the asymptotic average 
multiplicity inside a gluon jet (Eq.\ (\ref{eq:dlamult})), $\bar l$, $\sigma$, $s$ and $k$
are respectively the position of the peak, the dispersion, 
the skewness and the kurtosis of the distribution. 
They are given by
\begin{eqnarray}
\bar\ell&=&K_1,\nonumber \\
\sigma&=&\sqrt{K_2},\nonumber \\
s&=&\frac{K_3}{\sigma^3},\nonumber \\
k&=&\frac{K_4}{\sigma^4},
\label{dGparfromKn}
\end{eqnarray}
where $K_n$ is the $n$th moment of $D_g$:
$$
K_n(Q^2)=\left(-\frac{d}{d\omega}\right)^n\ln [D_g(\omega,Q^2)]_{\omega=0}.
$$
The reason why Eq.\ (\ref{dg}) is valid around the position of the peak (i.e.\ for $\delta=O(1)$)
is because the exponent is believed to be an expansion in $a_s$, and this is the case in the MLLA.
We now determine the distorted Gaussian parameters as a function of $a_s$ and also $N_s$.
The extension to the MLLA is given in appendix \ref{MLLAlim}.
According to Eq.\ (\ref{eq:solDg}), the $K_n(Q^2)$ evolve as
\begin{equation}
K_n(Q^2)=\Delta K_n(a_s(Q^2),a_s(Q_0^2))+K_n(Q_0^2),
\label{evolofKn}
\end{equation}
where $\Delta K_n(a_s(Q^2),a_s(Q_0^2))$ is the $n$th moment
of $E_{gg}(\omega,a_s(Q^2),a_s(Q_0^2))$, given by
$$
\Delta K_n(a_s(Q^2),a_s(Q_0^2))=\int_{Q_0}^Qd\ln Q'^2
\left(-\frac{d}{d\omega}\right)^nP_{gg}(\omega,a_s(Q'^2)).
$$
We now make the approximation that any constants and $Q_0$-dependent terms in Eq.\ (\ref{evolofKn}) can be neglected, which 
is valid for sufficiently large $Q$.
Making use of Eq.\ (\ref{eq:anomdim}), one finds
\begin{eqnarray}\label{eq:K}
K_1&=&\frac{a_s^{-1}}{4\beta_0},\nonumber \\
K_2&=&\frac{a_s^{-3/2}}{6\beta_0\sqrt{16C_AN_s}},
\nonumber \\
K_3&=&0,\nonumber \\
K_4&=&-\frac{3a_s^{-5/2}}{10\beta_0(16C_AN_s)^{3/2}}.
\end{eqnarray}
Then from Eqs.\ (\ref{dGparfromKn}) and (\ref{eq:K}), one finds
\begin{subequations}
\begin{eqnarray}
\bar\ell&=&\frac{a_s^{-1}}{4\beta_0},\label{eq:xi}\\
\sigma&=&\frac{a_s^{-3/4}}
{\sqrt{6\beta_0}(16C_AN_s)^{1/4}},\label{eq:sigma}\\
s&=&0,\label{eq:skew}\\
k&=&
-\frac{54}{5}\beta_0\sqrt{\frac{a_s}{16C_AN_s}}.\label{eq:kurt}
\end{eqnarray}
\end{subequations}
From these results, we find that the distribution around the position 
of the peak (i.e.\ $\delta=O(1)$) can be approximated by a distorted Gaussian,
namely Eq.\ (\ref{eq:smallxD}).
The width (Eq.\ (\ref{eq:sigma})) and 
the kurtosis (Eq.\ (\ref{eq:kurt})) of this distorted Gaussian are suppressed in the limit $N_s\gg1$ by 
$1/N_s^{\alpha}$, where $\alpha$ is a positive constant.

Setting $\omega=0$ in Eq.\ (\ref{eq:anomdim}), 
one recovers the rate of the mean average 
multiplicity increase in a gluon jet with respect to $\ln Q^2$, 
i.e.\ the DLA medium-modified anomalous 
dimension of the multiplicity \cite{Redmed}:
\begin{equation}\label{Patomegaeq0}
P_{gg}(0,a_s)=\sqrt{C_AN_sa_s},
\end{equation}
in which case Eqs.\ (\ref{eq:solDg}) and (\ref{eq:solDgbis}) give Eq.\ (\ref{eq:dlamult}) at large $Q$.

\section{Conclusions}

\label{section:conclusions}

Our approach provides a general framework for the incorporation of medium effects 
into the DGLAP evolution of FFs from large to small $x$,
which will be important for the description of single hadron production in heavy-ion
collisions from high to low $p_T$.
In particular, our approach allows the use of measurements of such observables at low $p_T$, e.g.\ from
heavy-ion collisions at the LHC, to provide additional constraints
on the medium-modified FFs to those constraints provided by large-$p_T$ data,
and therefore could improve the ruling out of various models of the medium
by improving the constraints on the unknown degrees of freedom.

\appendix

\section{MLLA limit}
\label{MLLAlim}

The MLLA can be regarded as a simplification to the approach of Ref.\ \cite{AKKPRL} and of this paper, in which the 
main qualitative features of hadron production are preserved.
We now use our approach to determine the form of the MLLA with medium effects taken into account.
The MLLA is obtained by setting $\omega=0$ in the second term in Eq.\ (\ref{DLdecofP2}),
i.e.\ by taking
\begin{equation}
P(\omega,a_s)=P^{\rm DL}(\omega,a_s)+a_s P^{\nDL (0)}(0),
\end{equation}
which for small $\omega$ is correct to terms of $O(\omega)$.
In the case of the medium,
\begin{equation}\label{eq:matelembis}
P^{\nDL (0)}(0)=\left(
\begin{array}{cc}
        2C_F f_{\Sigma \Sigma}(0) & 4C_Ff_{\Sigma g}(0)  \cr
        n_fT_Rf_{g\Sigma}(0) &  2C_A f_{gg}(0)
\end{array}\right),
\end{equation} 
which corresponds to the hard SL 
corrections in the MLLA formalism \cite{Basics}. Setting 
$$
D_{\Sigma}=\frac{2C_F}{C_A}D_g,
$$
the system in Eq.\ (\ref{eq:mellineq}) with Eq.\ (\ref{eq:matelembis}) 
reduces to a self-contained equation for $D_g$ and a coupled
equation for $D_{\Sigma}$, constrained by the solution of $D_g$,
which can be written in the form

\vbox{
\begin{eqnarray}\label{eq:qjet}
\left(\!\!\omega+2\frac{d}{d\ln Q^2}\!\!\right)\!
\frac{d}{d\ln Q^2}D_{\Sigma}\!\!&\!\!=\!\!&\!\!4C_FN_sa_sD_g\!\!+\!\!
\left(\!4C_Ff_{\Sigma g}(0)\!+\!\frac{2C_F}{C_A} f_{\Sigma \Sigma}(0)\!\right)
\!\!\left(\!\!\omega+2\frac{d}{d\ln Q^2\!\!}\right)\!\!a_sD_g,\\\notag\\
\left(\!\!\omega+2\frac{d}{d\ln Q^2}\!\!\right)\!
\frac{d}{d\ln Q^2}D_g\!\!&\!\!=\!\!&\!\!2C_AN_sa_sD_g-\frac{1}{2}
a_{\text{med}}\left(\!\!\omega+2\frac{d}{d\ln Q^2}\!\!\right)a_sD_g.
\label{eq:gjet}
\end{eqnarray}
}
In Ref.\ \cite{Urs}
the choice $N_s=1$ is made in $P^{\nDL (0)}(0)$, which is known:
\begin{equation}\label{eq:matelem}
[P^{\nDL (0)}(0)]_{N_s =1}=\left(
\begin{array}{cc}
        0 & -3C_F  \cr
        \frac23n_fT_R & -\frac{11}{6}C_A-\frac23n_fT_R 
\end{array}\right),
\end{equation}
and in $a_{\text{med}}$, which is also known:
$$
[a_{\text{med}}]_{N_s =1}=a=\frac{11C_A}{3}+\frac{4n_fT_R}{3}\left(1-\frac{2C_F}{C_A}\right).
$$
However, we will keep $N_s \neq 1$ in what follows in order to maintain generality.
Using Eqs.\ (\ref{eq:solDg}) and (\ref{eq:solDgbis}),
Eq.\ (\ref{eq:gjet}) after some straightforward algebraic operations becomes
\footnote{A term $+a_{\text{med}}\beta(a_s)$ on the right hand side has been neglected since it is not required in what follows.}
\begin{equation}
(\omega+2P_{gg})P_{gg}-2C_AN_sa_s=-2\beta(a_s)\frac{dP_{gg}}{da_s}-
\frac12a_{\text{med}}(\omega+2P_{gg})a_s,
\end{equation}
which provides the following solution correct to terms of single logarithmic order:
\begin{eqnarray}\label{eq:anomdimMLLA}
P_{gg}(\omega,a_s)\!\!&\!\!=\!\!&\!\!\frac14\left(-\omega+\sqrt{\omega^2+16C_AN_sa_s}\right)\\
\!\!&\!\!+\!\!&\!\!a_s\left[\beta_0\frac{4C_AN_sa_s}{\omega^2+16C_AN_sa_s}-
\frac14a_{\text{med}}
\left(1+\frac{\omega}{\sqrt{\omega^2+16C_AN_sa_s}}\right)\right].\notag
\end{eqnarray}
Setting $\omega=0$ in Eq.\ (\ref{eq:anomdimMLLA}), 
one recovers the rate of the mean average 
multiplicity increase in a gluon jet with respect to $\ln Q^2$, 
i.e.\ the MLLA medium-modified anomalous 
dimension of the multiplicity \cite{Redmed}:
\begin{equation}\label{Pprimeatomegaeq0}
P_{gg}(0,a_s)=\sqrt{C_AN_sa_s}+\frac{a_s}{4}
\left(\beta_0-a_{\text{med}}\right).
\end{equation}
The purpose of $\beta_0$
is to account for the running of the coupling constant $a_s$.
Finally, the solution for $D_\Sigma$ at NLO in Mellin space
can be derived from Eq.\ (\ref{eq:qjet}), once $D_g$
has been computed, and reads
\begin{equation}\label{eq:DFDg}
D_\Sigma=\frac{2C_F}{C_A}\left\{1+\frac1{N_s}\left[
\frac1{4C_A}a_{\text{med}}+f_{\Sigma g}(0)+
\frac{C_F}{C_A}f_{\Sigma \Sigma}(0)\right]\left(\omega+2\frac{d}{d\ln Q^2}\right)\right\}D_g.
\end{equation}

Since $\left(\omega+2\frac{d}{d\ln Q^2}\right)\simeq{\cal O}(\sqrt{N_s a_s})$, 
the second term on the right-hand side of Eq.\ (\ref{eq:DFDg}) is of order ${\cal O}(\sqrt{a_s/N_s})$.
For $N_s=1$, one recovers the result in the vacuum \cite{MALAZA} with
$$
\left[\frac1{4C_A}a_{\text{med}}+f_{\Sigma g}(0)+
\frac{C_F}{C_A}f_{\Sigma \Sigma}(0)\right]_{N_s=1}=
\frac{a}{4C_A}-\frac34.
$$

Making use of Eq.\ (\ref{eq:anomdimMLLA}), one finds
\begin{eqnarray}\label{eq:KMLLA}
K_1&=&\frac{a_s^{-1}}{4\beta_0}+\frac{a_{\text{med}}a_s^{-1/2}}{2\beta_0\sqrt{16C_AN_s}},\nonumber \\
K_2&=&\frac{a_s^{-3/2}}{6\beta_0\sqrt{16C_AN_s}}
-\frac{a_s^{-1}}{32C_AN_s},\nonumber \\
K_3&=&-\frac{a_{\text{med}}a_s^{-3/2}}{2\beta_0(16C_AN_s)^{3/2}},\nonumber \\
K_4&=&-\frac{3a_s^{-5/2}}{10\beta_0(16C_AN_s)^{3/2}}+
\frac{3a_s^{-2}}{(16C_AN_s)^2}.
\end{eqnarray}
Then from Eqs.\ (\ref{dGparfromKn}) and (\ref{eq:KMLLA}), one finds
\begin{subequations}
\begin{eqnarray}
{\cal N}&=&{\cal N}_0 \exp \left[\frac{2}{\beta_0}\sqrt{\frac{C_A N_s}{a_s}}+\frac{1}{4\beta_0}\left(a_{\rm med}-\beta_0\right)
\ln a_s\right]\\
\bar\ell&=&\frac{a_s^{-1}}{4\beta_0}+\frac{a_{\text{med}}
a_s^{-1/2}}{2\beta_0\sqrt{16C_AN_s}},\label{eq:xiMLLA}\\
\sigma&=&\frac{a_s^{-3/4}}
{\sqrt{6\beta_0}(16C_AN_s)^{1/4}}
\left(1-\frac32\beta_0\sqrt{\frac{a_s}{16C_AN_s}}\right),\label{eq:sigmaMLLA}\\
s&=&-3\sqrt{6\beta_0}\,a_{\text{med}}
\left(\frac{a_s}{16C_AN_s}\right)^{3/4},\label{eq:skewMLLA}\\
k&=&
-\frac{54}{5}\beta_0\sqrt{\frac{a_s}{16C_AN_s}}
\left(1-4\beta_0\sqrt{\frac{a_s}{16C_AN_s}}\right),\label{eq:kurtMLLA}
\end{eqnarray}
\end{subequations}
which coincide with the results in Ref.\ \cite{fonweb} in the limit $N_s =1$.

Assuming that $a_{\text{med}}$ rises slower than $N_s^{3/4}$, the skewness (Eq.\ (\ref{eq:skewMLLA})), 
like the width (Eq.\ (\ref{eq:sigmaMLLA}))
and the kurtosis (Eq.\ (\ref{eq:kurtMLLA})), is suppressed in the limit $N_s\gg1$.
As found in subsection \ref{smallx}, the distribution close to the position 
of the peak can be approximated by a Gaussian shape.
Finally, in Ref.\ \cite{Urs}, the choice $\left[a_{\text{med}}\right]_{N_s=1}=a$ is made, 
in which case the position of the peak approaches the asymptotic DLA value 
$\frac{a_s^{-1}}{4\beta_0}$ for large $N_s$.

%%%%%%%%%%%%%%%%%%%%%%%%%%%%%%%%%%%%%%%%%%%%%%%%%%%%%%%%%%%%%%%%%%%%%%%%%%%%%%
%%%%%%%%%%%%%%%%%%%%%%%%%%%%%%%%%%%%%%%%%%%%%%%%%%%%%%%%%%%%%%%%%%%%%%%%%%%%%%
\null

%%%%%%%%%%%%%%%%%%%%%%%%%%%%%%%%%%%%%%%%%%%%%%%%%%%%%%%%%%%%%%%%%%%%%%%%%%%

\end{document}